\documentclass{article}

\usepackage{arxiv}

\usepackage[utf8]{inputenc} 
\usepackage[T1]{fontenc}    
\usepackage{hyperref}       
\usepackage{url}            
\usepackage{booktabs}       
\usepackage{amsfonts}       
\usepackage{nicefrac}       
\usepackage{microtype}      
\usepackage{lipsum}
\usepackage{graphicx}
\graphicspath{ {./images/} }

\title{A Novel Approach to Image EEG Sleep Data for Improving Quality of Life in Patients Suffering From Brain Injuries Using DreamDiffusion}

\author{
 David Fahim \\
   \And
 Joshveer Grewal \\
   \And
 Ritvik Ellendula \\
}

\begin{document}
\maketitle

\vspace{\baselineskip}
\begin{abstract}
Those experiencing strokes, traumatic brain injuries, and drug complications can often end up hospitalized and diagnosed with coma or locked-in syndrome. Such mental impediments can permanently alter the neurological pathways in work and significantly decrease the quality of life (QoL). It is critical to translate brain signals into images to gain a deeper understanding of the thoughts of a comatose patient. Traditionally, brain signals collected by an EEG could only be translated into text, but with the novel method of an open-source model available on GitHub, DreamDiffusion can be used to convert brain waves into images directly. DreamDiffusion works by extracting features from EEG signals and then using the features to create images through StableDiffusion. Upon this, we made further improvements that could make StableDiffusion the forerunner technology in waves to media translation. In our study, we begin by modifying the existing DreamDiffusion codebase so that it does not require any prior setup, avoiding any confusing steps needed to run the model from GitHub. For many researchers, the incomplete setup process, errors in the existing code, and a lack of directions made it nearly impossible to run, not even considering the model's performance. We brought the code into Google Colab so users could run and evaluate problems cell-by-cell, eliminating the specific file and repository dependencies. We also provided the original training data file so users do not need to purchase the necessary computing power to train the model from the given dataset. The second change is utilizing the mutability of the code and optimizing the model so it can be used to generate images from other given inputs, such as sleep data. Additionally, the affordability of EEG technology allows for global dissemination and creates the opportunity for those who want to work on the shared DreamDiffusion model.
\end{abstract}


\section{Introduction}
Our project focused on creating a method to facilitate communication with those impacted by brain impairments, utilizing EEG waves as the primary tool. Therefore, our initiative differs from other treatment strategies, as we strive to create a non-invasive method to facilitate communication rather than directly providing them with drugs and other medicines to impede the development of the brain ailment. To do so, we expanded off existing methodologies, such as DreamDiffusion and StableDiffusion, modifying and cleaning up the code segments to create an open-source method that uses the EEG data to predict and image those thoughts. Treatment modalities for brain injuries constitute interventions tailored to address the vast array of physical, emotional, and cognitive ramifications associated with the condition. In instances of mild TBI, colloquially referred to as concussions, therapeutic protocols predominantly emphasize rest, complemented by adherence to healthcare provider directives aimed at minimizing symptom exacerbation and facilitating optimal recovery. Activities demanding cognitive exertion, such as prolonged computer usage, are typically curtailed. At the same time, abstention from alcohol and drugs is strongly advocated to forestall potential setbacks in rehabilitation and mitigate the risk of re-injury, which could precipitate enduring neurological deficits or even mortality. Urgent management of TBI is centered on expeditiously stabilizing patients, ensuring adequate cerebral perfusion and oxygenation, often necessitating surgical intervention to address complications such as intracranial hemorrhage, skull fractures, or elevated intracranial pressure. Pharmacological interventions have a pivotal role in symptom amelioration and prophylaxis, encompassing a spectrum of medications, including anxiolytics, anticoagulants, anticonvulsants, antidepressants, diuretics, muscle relaxants, and stimulants, tailored to individual symptomatology and therapeutic difficulties. Additionally, rehabilitative strategies constitute a cornerstone of TBI management, employing an interdisciplinary approach comprising physical therapy, occupational therapy, speech-language pathology, psychological counseling, vocational rehabilitation, and cognitive remediation to optimize functional recovery, promote independence, and enhance quality of life. These rehabilitative endeavors are typically administered across a continuum of care settings, ranging from acute hospitalization to community-based outpatient services, with treatment plans meticulously individualized to accommodate patient-specific needs, prognostic factors, and long-term functional objectives [1]. Even though all these treatment types aid in marginal improvements, none can track the extent of recovery.

\section{Etiology}
\label{sec:headings}
It is worth mentioning that TBIs are a broad category of injuries that occur in the brain, affecting the same set of locations. The major types are concussions, contusions, tSAH, and hematoma. A concussion denotes a mild head injury typically characterized by a brief loss of consciousness, often devoid of enduring cerebral damage. A contusion signifies localized brain bruising resulting from head impact, alternatively termed a coup or contrecoup injury. In coup instances, brain trauma occurs directly beneath the point of impact, whereas in contrecoup scenarios, injury manifests on the opposite side of the effects. Traumatic Subarachnoid Hemorrhage (tSAH) describes bleeding into the cerebral space enveloping the brain. Generally filled with cerebrospinal fluid (CSF), this space is a protective cushion for the brain. Traumatic SAH arises from the tearing of small arteries during the initial injury, leading to blood dispersion across the brain's surface and subsequent wide-ranging effects. Hematoma refers to a blood clot formation following blood vessel rupture. Escaping blood thickens and coagulates outside the regular circulatory system. Coagulation serves as the body's innate mechanism to staunch bleeding. Hematomas may vary in size and exert pressure on the brain, with symptoms contingent upon the clot's location. An epidural hematoma emerges between the skull and the brain's dura lining, while a subdural hematoma forms between the brain and the dura. An intracerebral hematoma develops deep within brain tissue. Eventually, the body reabsorbs the clot, and surgical intervention is sometimes necessary for more considerable clot removal [2]. From this point on, a reference to TBIs includes any of these injuries that demonstrate a similar degree of trauma to the brain.

\noindent As there are many causes of TBIs among different populations with dissimilar groups in relationship with age, race, and social determinants, the affected brain region varies. Each region controls functions associated with it, yielding a wide range of symptoms that have proved troublesome to treat. The brain is divided into four major lobes and two brain stem sections (Figure 1).
\begin{figure}[htbp]
  \centering
  \includegraphics[width=0.5\textwidth]{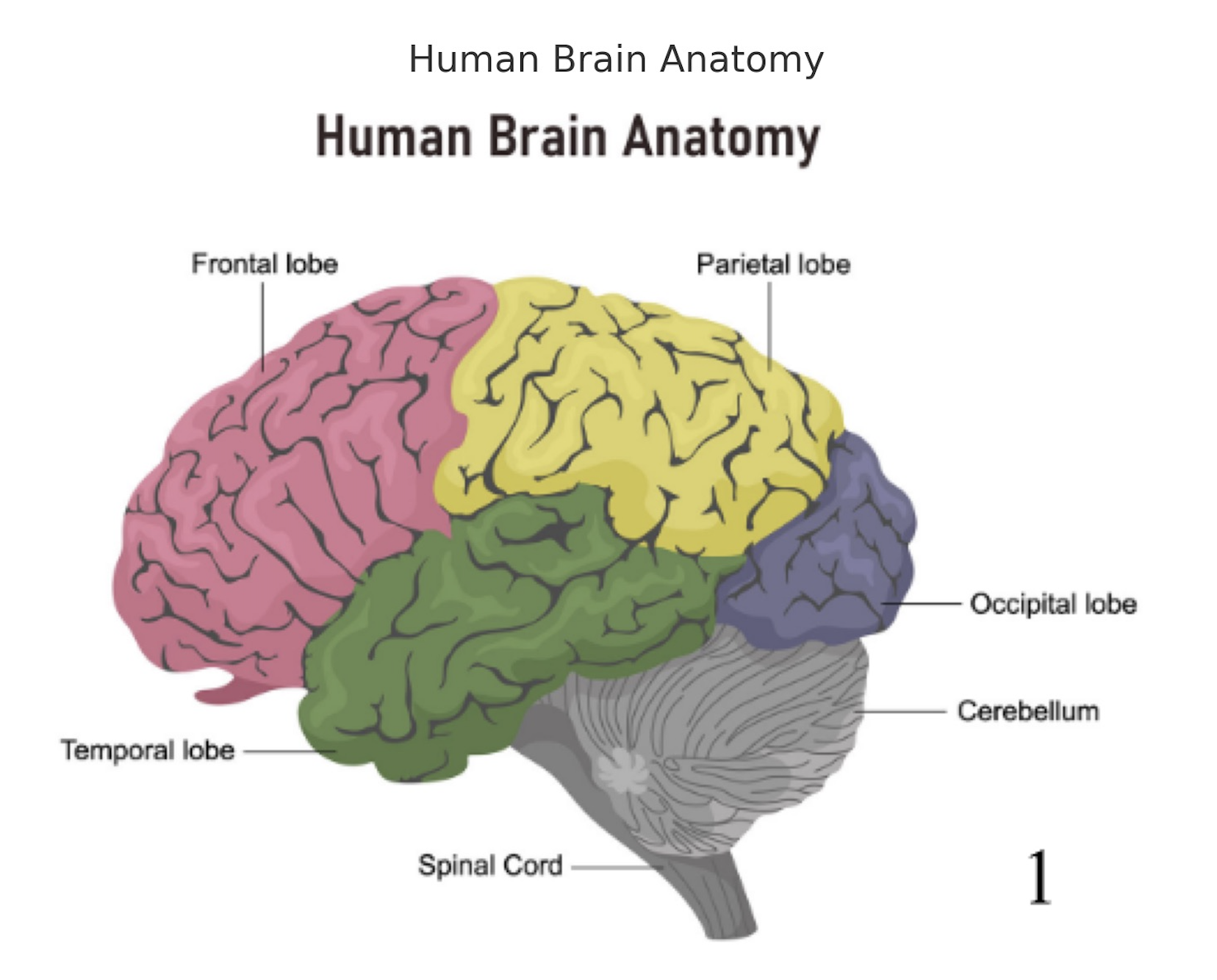}
  \caption{Shows the different regions of the brain by coloration. This includes the frontal, parietal, temporal, and occipital lobes, as well as the cerebellum and brainstem (spinal cord) [13].}
  \label{fig:image}
\end{figure}

\def\myvar{defined}

\noindent A simple breakdown of each location’s altered function provides the context of what aspect of human nature they control (Figure 2).

\begin{figure}[htbp]
  \centering
  \includegraphics[width=0.5\textwidth]{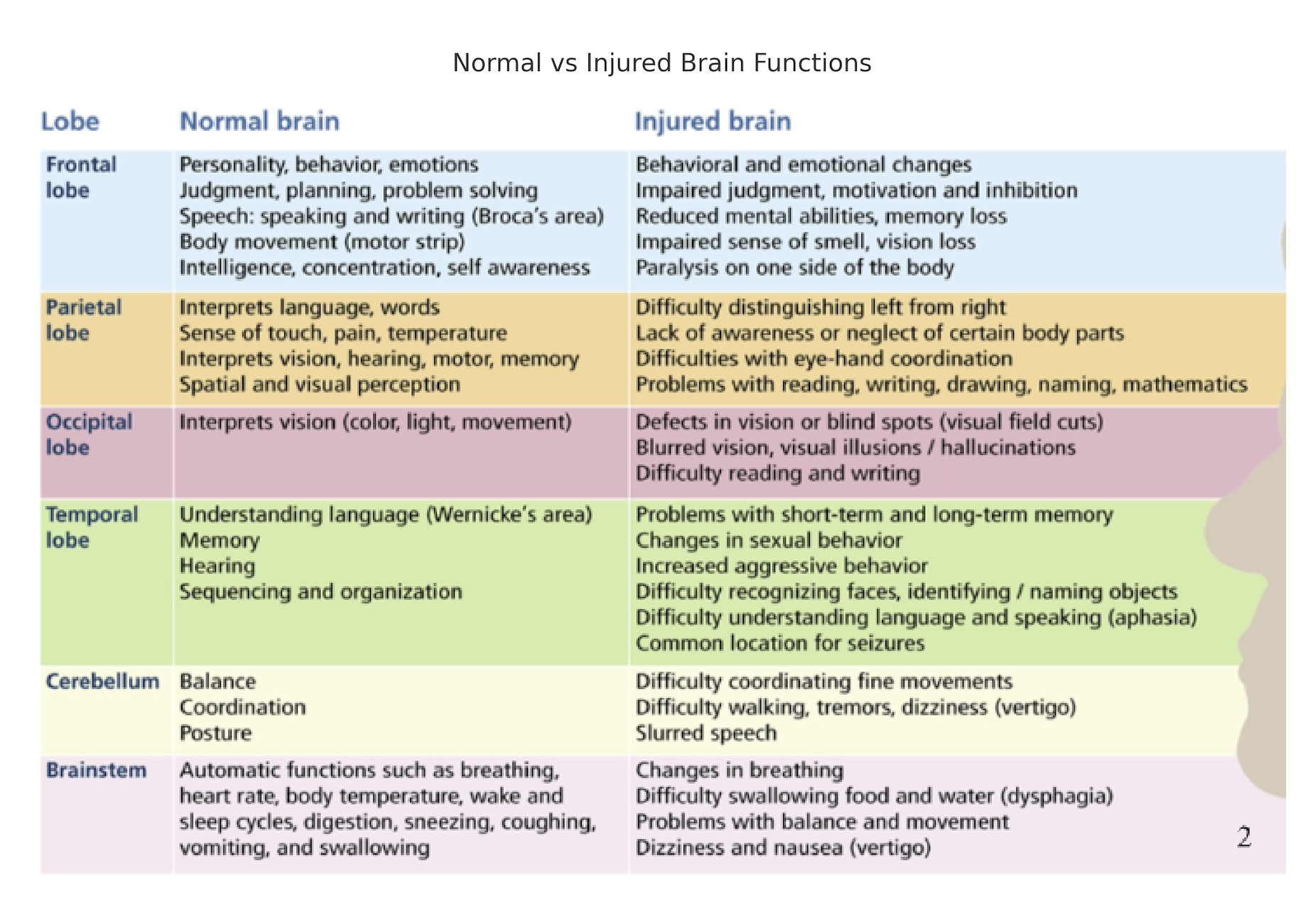}
  \caption{Shows the brain's normal and impaired function of various regions commonly associated with TBIs [14].}
  \label{fig:image}
\end{figure}
\def\myvar{defined}

\noindent Although the behaviors for each region can be outlined, there is much confounding on conditions resulting in incomplete or incorrect diagnosis. 35.1 percent of brain injuries were more likely to be diagnosed as TBIs. The study calculated an odds ratio (OR) of 1.351, which means that somebody with a brain injury is 1.351 times more likely to have it diagnosed as a TBI than somebody who does not [3]. Also, it is imperative to note that a binomial logistical regression model was used to examine the association between TBI and the gold standard, yielding the highest R2 value (Figure 3). The model was adjusted for age at death, sex, and race and returned various p-values, each significant to the ubiquitous alpha level of 0.05. Age at death and sex were the most common biases for diagnosing the disease as a TBI, both with p-values of  <0.0001 (Figure 4).

\begin{figure}[htbp]
  \centering
  \includegraphics[width=0.5\textwidth]{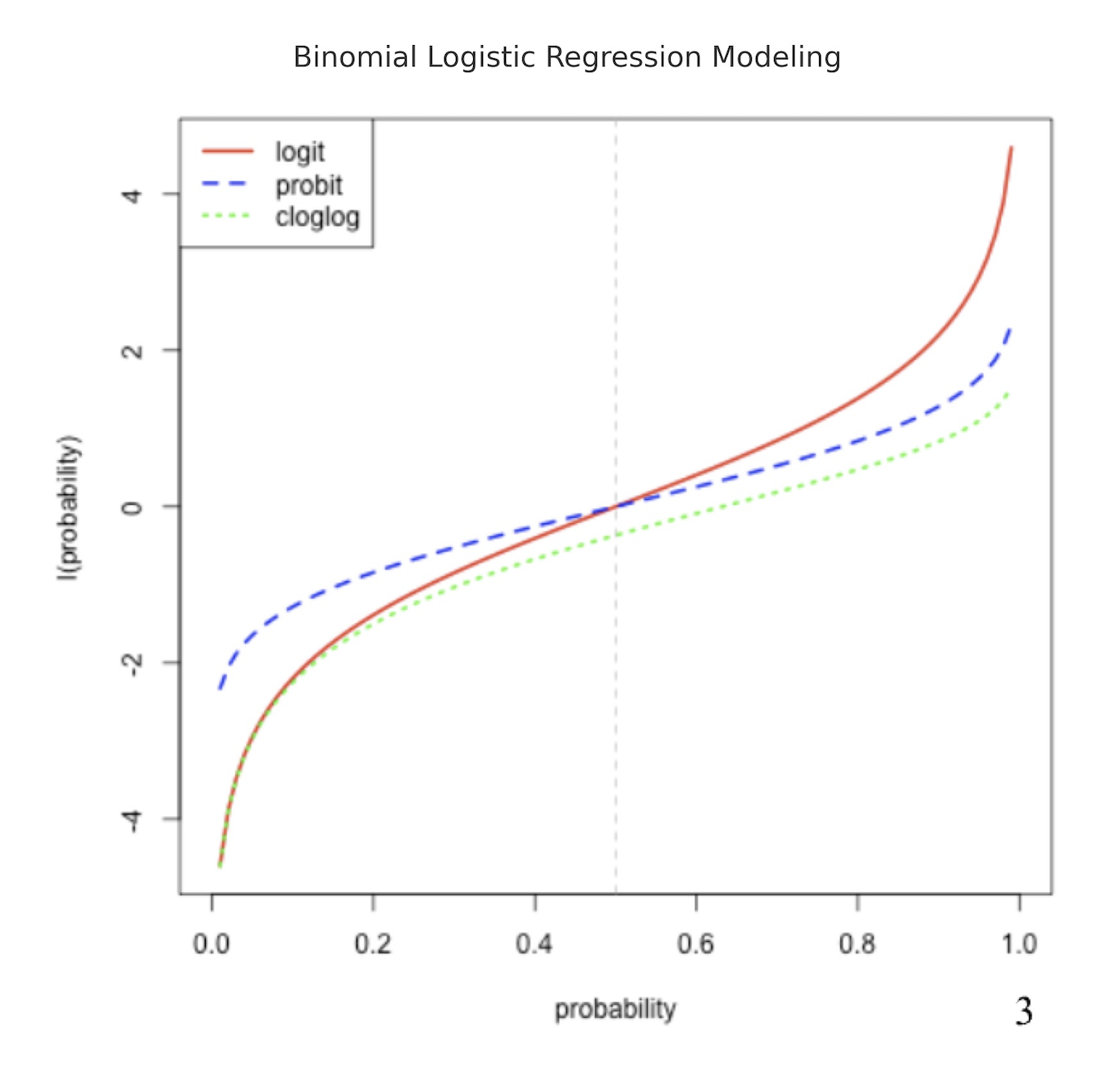}
  \caption{Shows the relationship between concavity and probability on a binomial logistical regression curve [15].}
  \label{fig:image}
\end{figure}
\def\myvar{defined}

\begin{figure}[htbp]
  \centering
  \includegraphics[width=0.5\textwidth]{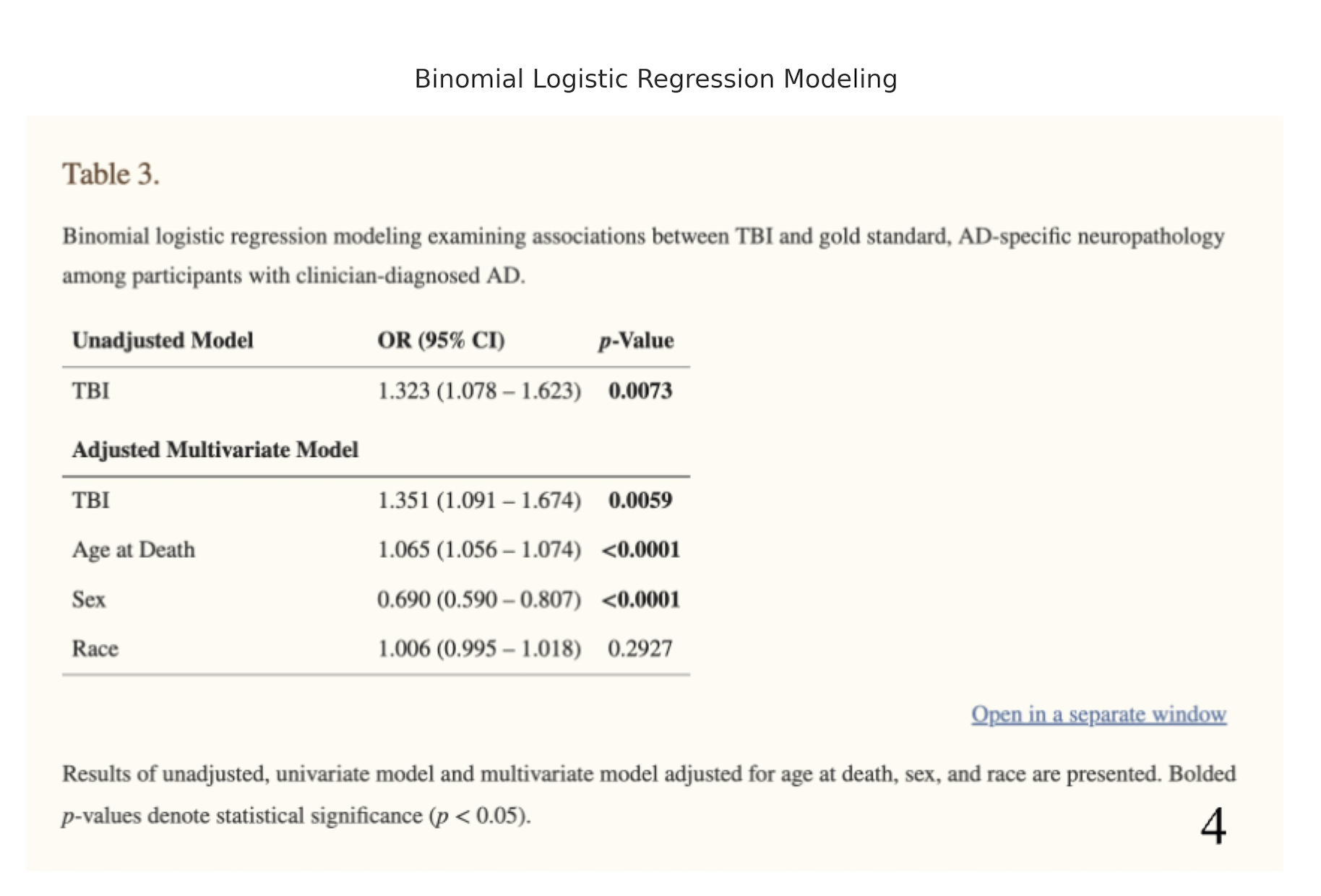}
  \caption{Shows the representation between normal and impaired function of the brain's various regions commonly associated with TBIs [16].}
  \label{fig:image}
\end{figure}
\def\myvar{defined}

\noindent In many cases, due to the combination of confounding effects of the multiple neurological pathways, patients go into a state of prolonged loss of consciousness, also known as a coma. Coma patients are often described as not being in a state of neither consciousness nor deep sleep, while some recent studies have shown that many still have brain waves. Rooting from TBIs, the level of consciousness is measured on the Glasgow scale ranging from 3-15, where 8 is the threshold for diagnosis [4]. In rehabilitation settings, the Rancho scale is used to quantify the amount of recovery made (Figure 5). Even with both scales, there is no definite way to track patients' recovery progress, as the conditions can range from weeks to years, leading to persistent vegetation. Although more apparent with comas, TBIs face the unfortunate dilemma of an incomplete tracking of the illness without any steps to improve quality of life (QoL). 

\begin{figure}[htbp]
  \centering
  \includegraphics[width=0.5\textwidth]{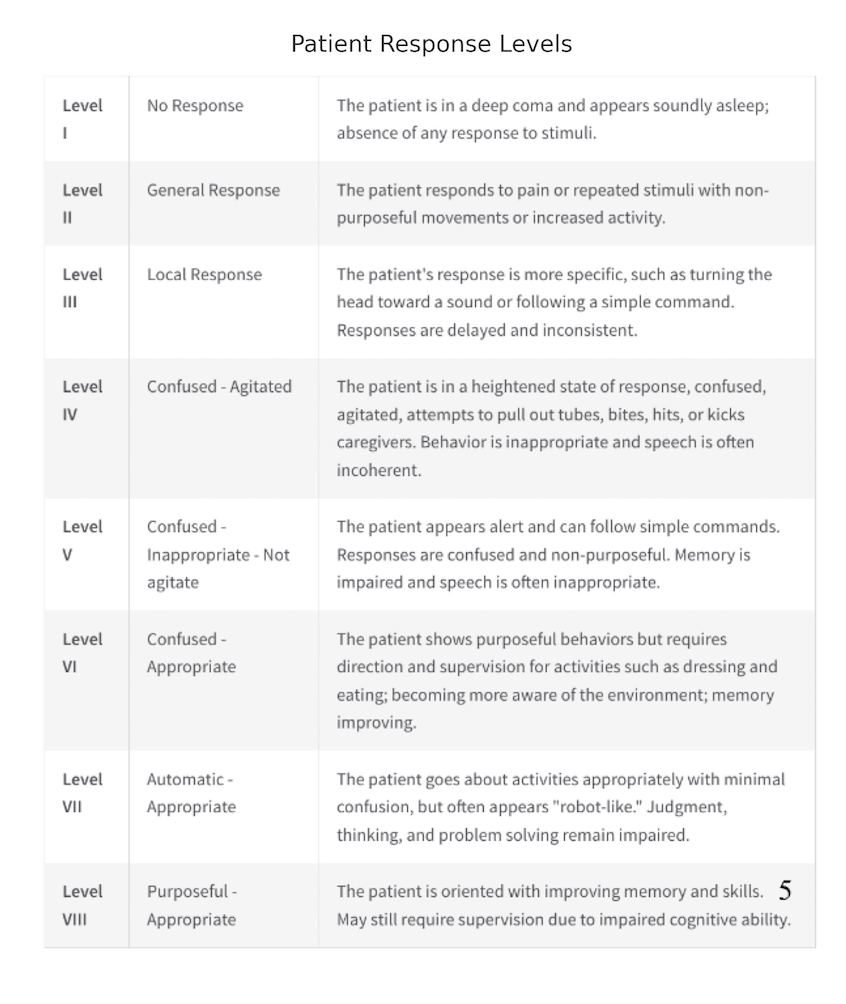}
  \caption{Illustrates the levels associated with various standards of consciousness on the Rancho scale [17].}
  \label{fig:image}
\end{figure}
\def\myvar{defined}

\section{Methodology}
\label{sec:headings}
\paragraph{EEG:}An electroencephalogram (EEG) utilizes electrodes attached to the scalp to capture electrical brain activity [5]. Comparable to other diagnostic tests such as fMRI and MRI, EEG can detect various conditions stemming from TBIs, albeit at different marginal costs. While fMRI and MRI yield superior-quality brain activity records with minimal noise, they entail longer acquisition times and significant equipment expenses. Conversely, EEG offers cost-effectiveness and quicker turnaround between scans, albeit with inherent signal noise. The process yields three states of the EEG signal: original, deconstructed, and reconstructed. Notice how the reconstructed signal's general trend closely corresponds with the original signal collected from patients (Figure 6). Specifically, one prominent oscillation observed is the gamma wave, commonly associated with cognitive processes such as thinking or problem-solving (Figure 7).

\noindent Developing an EEG Encoder that can encode EEG signals into semantic representations that can be accurately decoded to match the overall EEG signal is paramount. Embeddings are tokens positioned in nth-dimensional space. Due to the limited EEG-image pair data, instead of immediately training to generate embeddings that accurately represent images, the DreamDiffusion EEG Encoder is first trained to generate embeddings that represent a variety of unique EEG signals. Therefore, we used a more substantial dataset of EEG recordings rather than a limited collection of EEG-image pairs. Segments of this EEG data are randomly obscured, leaving behind a masked EEG signal that is then tokenized and fed into the EEG Encoder. The encoder generates embeddings, which are then passed to a decoder that reconstructs the masked portions of the signal. The overall trends of the reconstructed and original signal are compared to assess the accuracy of the EEG Encoder's embedding space.

\begin{figure}[htbp]
  \centering
  \includegraphics[width = 1\textwidth]{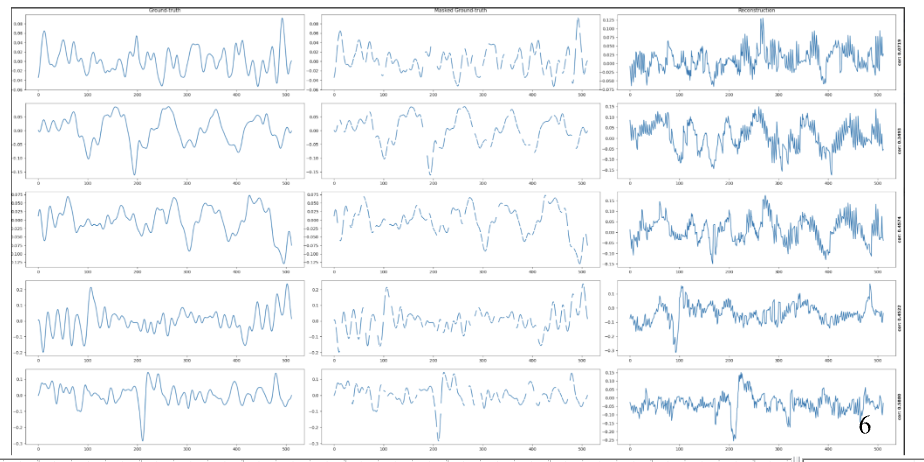}
  \caption{Shows the steps required to remove any unnecessary noise from the EEG wave data [18].}
  \label{fig:image}
\end{figure}
\def\myvar{defined}

\begin{figure}[htbp]
  \centering
  \includegraphics[width = 1\textwidth]{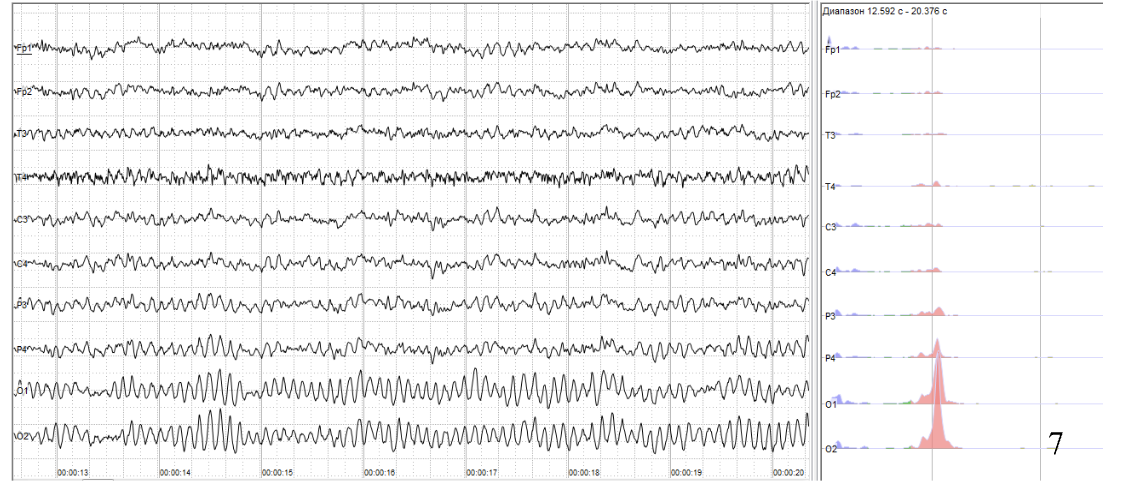}
  \caption{Portrays various gamma waves [19].}
  \label{fig:image}
\end{figure}
\def\myvar{defined}

\paragraph{CLIP:}Contrastive Language-Image Pre-training (CLIP), a simplified version of ConVIRT (Contrastive Learning of Medical Visual Representations from Paired Images and Text) trained from scratch, is an efficient means of learning image representations through natural language guidance. CLIP conducts joint training of an image encoder and a text encoder to accurately predict the pairings within a batch of (image, text) training examples. During testing, the learned text encoder constructs a zero-shot linear classifier by embedding the names or descriptions of the target dataset’s classes. During pre-training, CLIP is tasked with discerning the actual occurrences among the potential NXN (image, text) pairings within a batch. It achieves this by cultivating a multi-modal embedding space, where the image and text encoders are trained concurrently to maximize the cosine similarity of embeddings for the batch's N2 - N real pairs while minimizing the similarity for incorrect pairings (Figure 8). This objective is optimized through symmetric cross-entropy loss over the similarity scores. With CLIP trained on a diverse assortment of (image, text) pairs, it can be instructed through natural language to predict the most relevant text snippet corresponding to an image without explicit optimization for the task, reminiscent of the zero-shot capabilities showcased by GPT-2 and GPT-3 [6].

\begin{figure}[htbp]
  \centering
  \includegraphics[width = 1\textwidth]{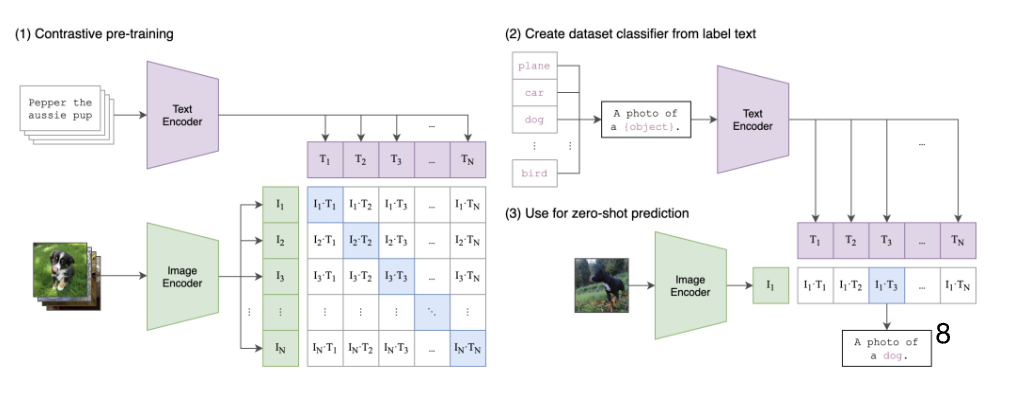}
  \caption{Shows the process of the text prompt creating the respective image [20].}
  \label{fig:image}
\end{figure}
\def\myvar{defined}

\paragraph{Stable Diffusion:}Stable Diffusion operates as a diffusion model, distinguishing it from conventional image generation models. Essentially, diffusion models utilize Gaussian noise to encode an image and subsequently employ a noise predictor with a reverse diffusion process to reconstruct the image. What sets Stable Diffusion apart is its departure from utilizing the pixel space of the image. Instead, it employs a reduced-definition latent space. This choice is motivated by the immense number of possible values in the pixel space of a color image with a resolution of 512x512, amounting to 786,432 values. In contrast, Stable Diffusion employs a compressed image with only 16,384 values, reducing processing demands significantly. The efficacy of the smaller latent space hinges on the non-random nature of natural images, which Stable Diffusion leverages by utilizing variational autoencoder (VAE) files in the decoder to capture fine details like eyes. The architectural components of Stable Diffusion include a VAE, forward and reverse diffusion, a noise predictor, and text conditioning. The VAE encompasses separate encoder and decoder components. The encoder compresses the 512x512 pixel image into a smaller 64x64 model in the latent space, facilitating easier manipulation. Conversely, the decoder restores the model from latent space into a full-size 512x512 pixel image. Forward diffusion systematically introduces Gaussian noise to an image until only random noise remains, rendering the final noisy image unidentifiable. All images undergo this process during training, although forward diffusion is only employed for image-to-image conversion. Reverse diffusion is a parameterized process that iteratively reverses the forward diffusion, allowing for the generation of unique images based on prompts and training data. The noise predictor, implemented as a U-Net model, is essential for denoising images. It estimates the noise in the latent space and subtracts it from the image, iteratively reducing noise according to user-specified steps. This component is sensitive to conditioning prompts that aid in determining the final image. Text conditioning, a common form, involves text prompts to guide image generation. Each word in a textual prompt is analyzed by a CLIP tokenizer and embedded into a 768-value vector. Stable Diffusion feeds these prompts from the text encoder to the U-Net noise predictor via a text transformer, enabling the generation of diverse images in the latent space by adjusting the random number generator's seed [7]. This process creates photorealistic images from text (Figure 9).

\begin{figure}[h]
  \centering
  \includegraphics[width = 0.7\textwidth]{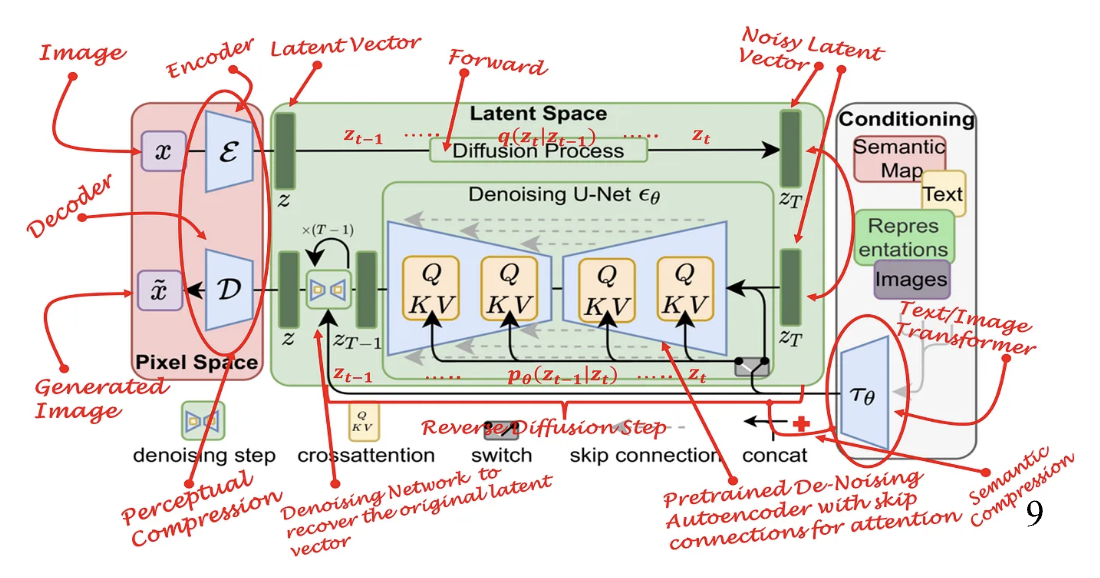}
  \caption{Illustrates the holistic labeled process of Stable Diffusion [21].}
  \label{fig:image}
\end{figure}
\def\myvar{defined}

\paragraph{DreamDiffusion:}The implementation of DreamDiffusion in Colab utilizes a Google Drive folder for storage, where various data, including images and EEG data, are downloaded by the Colab notebook. DreamDiffusion integrates Stable Diffusion, which employs an encoder called CLIP to convert text into embeddings for image generation. However, DreamDiffusion aims to replace CLIP with its own EEG Signals Encoder while still utilizing Stable Diffusion's image generator. The objective is to align the embedding space of the DreamDiffusion EEG Encoder with CLIP's embedding space. Given the inherent noise in EEG data, DreamDiffusion employs a strategy to address this challenge. Initially, it trains on a large dataset of EEG data rather than a limited collection of EEG-image pairs. Portions of this EEG data are randomly masked, and the remaining signal is tokenized before being fed into the encoder, which yields embeddings. Subsequently, a decoder model is employed to generate the masked parts in the signal from the embeddings, and the reconstructed signal is compared against the original. However, the embedding space produced by DreamDiffusion differs significantly from CLIP's embedding space, but CLIP's embedding space is ideal because it closely corresponds to images. To bridge this gap, a smaller dataset of EEG-image pairs is utilized. The image is encoded into CLIP embeddings for each pair to obtain the appropriate embeddings for Stable Diffusion. Simultaneously, the EEG signal is fed into the previously trained EEG Encoder, generating its embedding [12]. A comparison is made between CLIP and the EEG Encoder embeddings, and DreamDiffusion adjusts the EEG Encoder embedding space to minimize any disparity between these embedding spaces (Figure 10).

\begin{figure}[h]
  \centering
  \includegraphics[width = 0.7\textwidth]{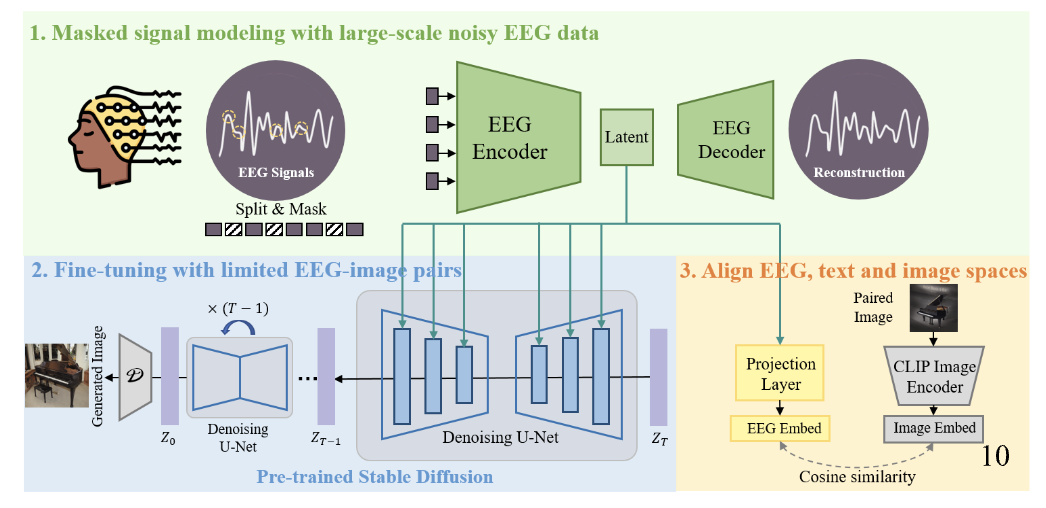}
  \caption{Shows the process of DreamDiffusion [22].}
  \label{fig:image}
\end{figure}
\def\myvar{defined}

\section{Results}
\label{sec:headings}
\paragraph{Training Data:} Before the EEG pretrain, we trained the model on the training data. On Colab Pro+, it took around 15 hours and got down to a loss of approximately 0.000243. This is calculated by finding the difference between 1 and the training loss step. Our code is shown below (Figures 11a, b, and c). 
Link to code: 

\url{https://colab.research.google.com/drive/1XwZb-IDR1_cxr47Y9qLD3Mvitq4HAuY5#scrollTo=3D5E6Oa0870-&uniqifier=3}

\begin{figure}[h]
  \centering
  \includegraphics[width = 0.8\textwidth]{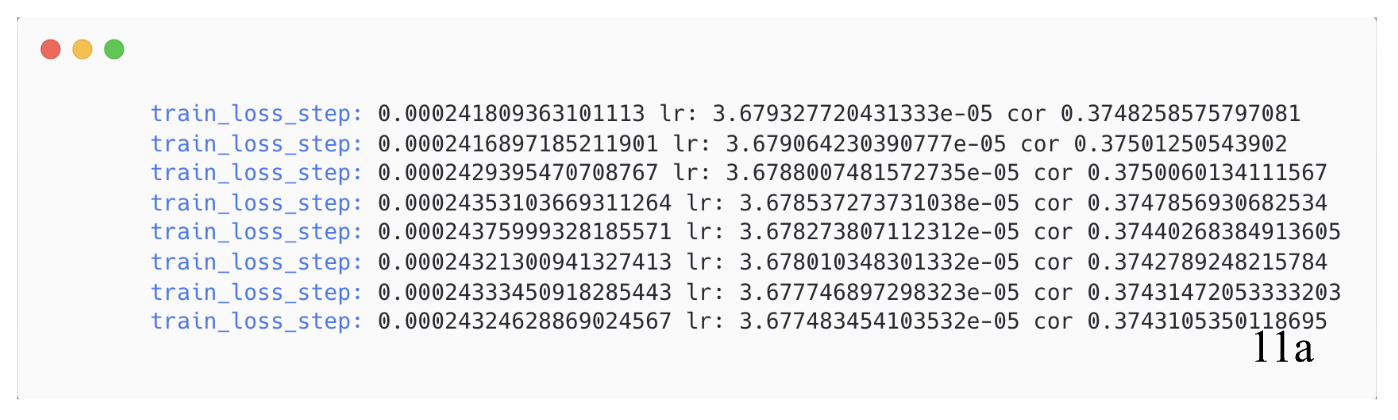}
  \label{fig:image}
\end{figure}
\def\myvar{defined}

\begin{figure}[h]
  \centering
  \includegraphics[width = 0.8\textwidth]{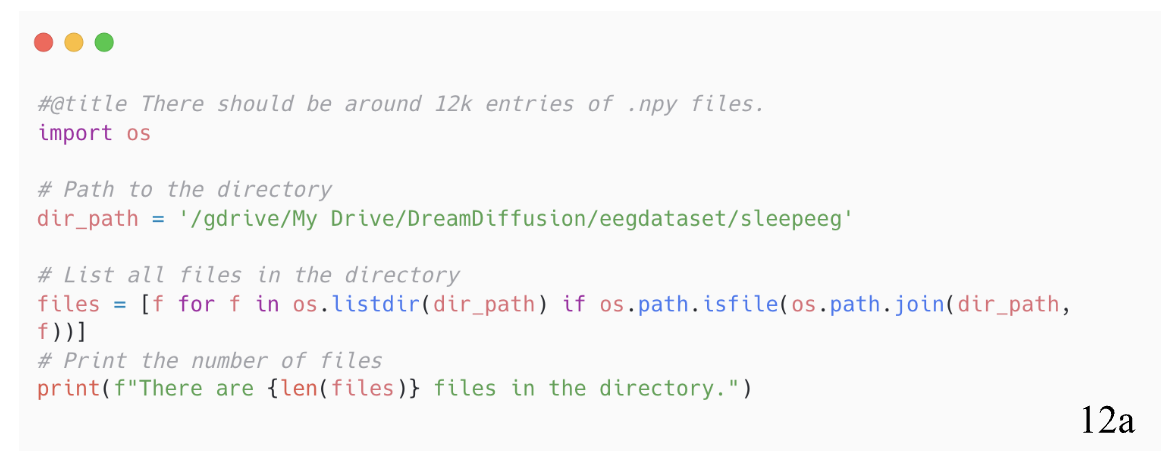}
  \label{fig:image}
\end{figure}
\def\myvar{defined}

\begin{figure}[h]
    \centering
    \begin{minipage}[b]{0.45\textwidth}
        \centering
        \includegraphics[width=\textwidth]{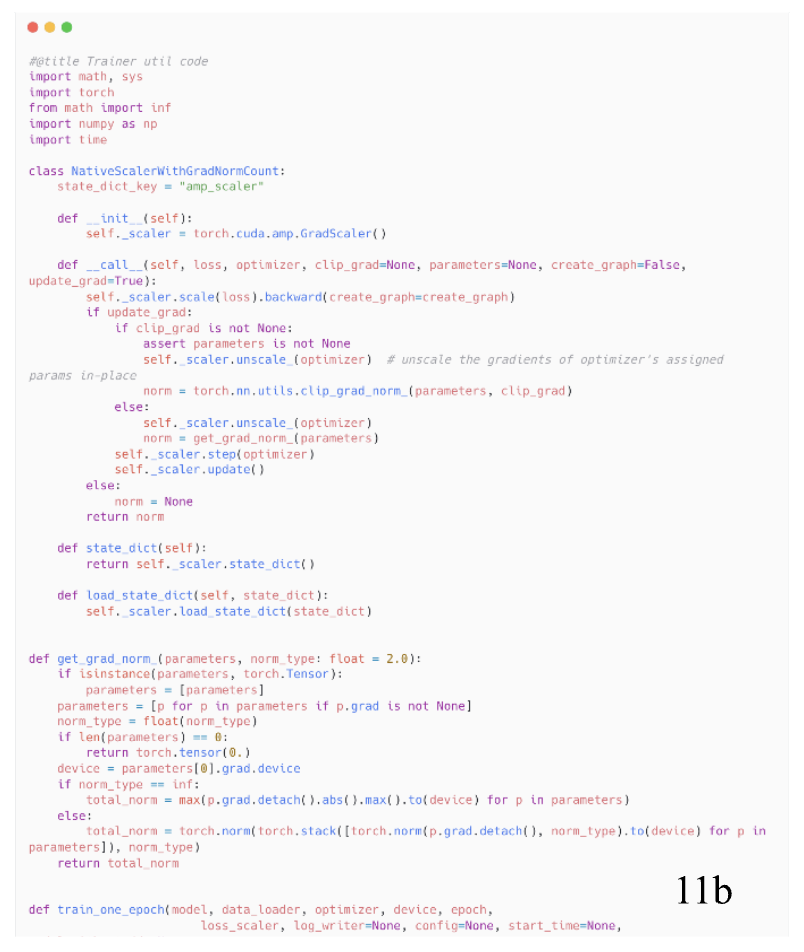}
    \end{minipage}
    \hspace{0.1cm} 
    \begin{minipage}[b]{0.45\textwidth}
        \centering
        \includegraphics[width=\textwidth]{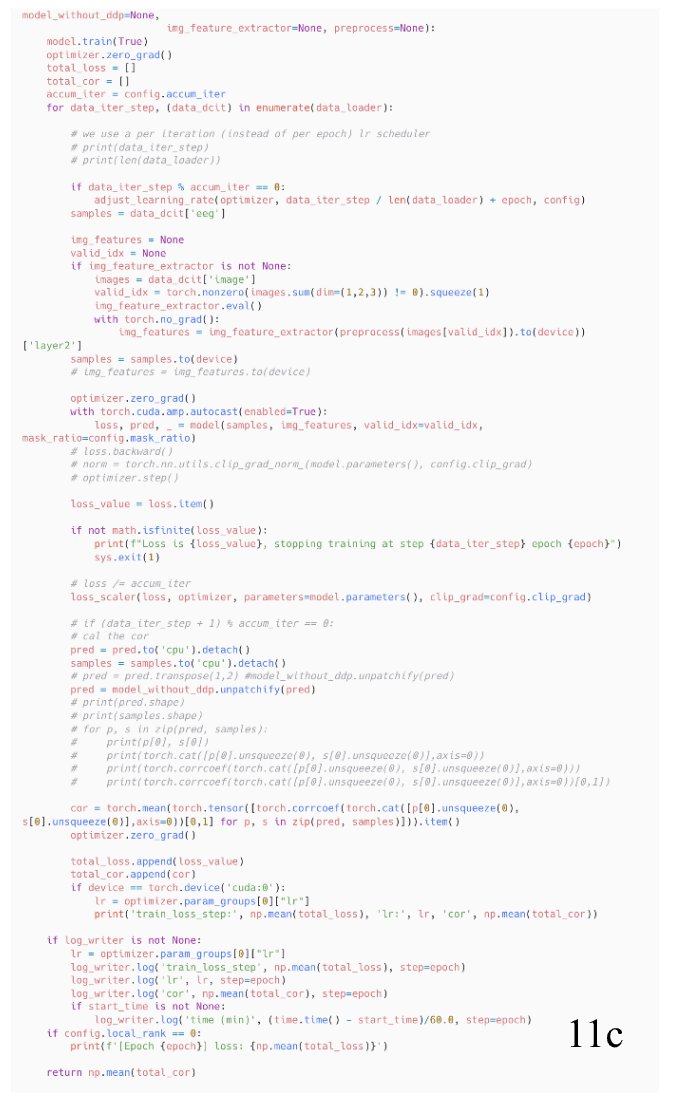}
    \end{minipage}
    \caption{11b and 11c show the training data code, while 11a depicts the training loss step [23].}
    \label{fig:side-by-side}
\end{figure}

\paragraph{EEG Input:}We imported EEG files and found that the image creation could accurately reflect the patients' dreams. Below are some reconstructed images that could be created from EEG waves (Figures 12a, b, and c).

\begin{figure}[htbp]
    \centering
    \begin{minipage}[b]{0.45\textwidth}
        \centering
        \includegraphics[width=\textwidth]{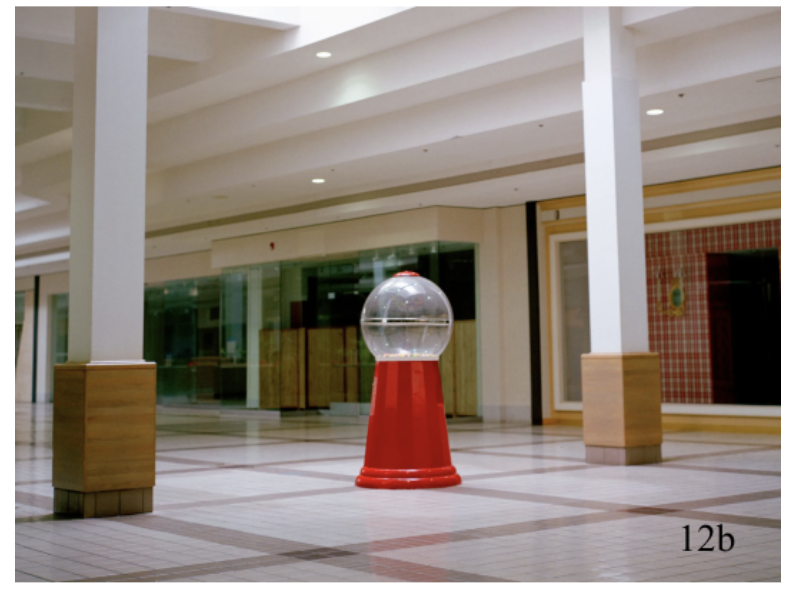}
    \end{minipage}
    \hspace{1cm} 
    \begin{minipage}[b]{0.45\textwidth}
        \centering
        \includegraphics[width=\textwidth]{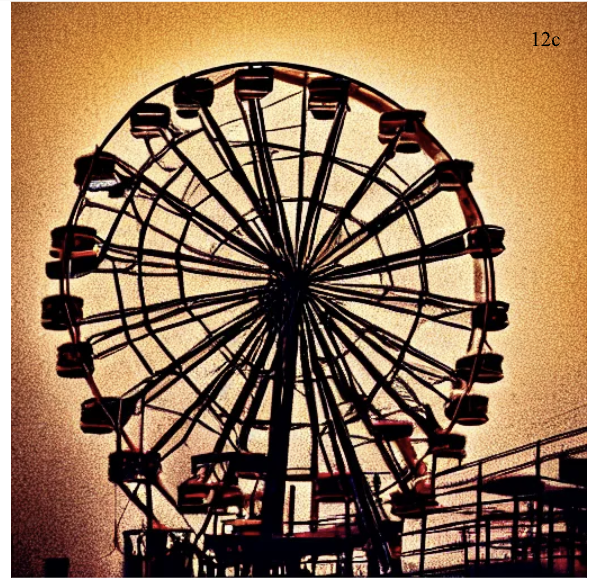}
    \end{minipage}
    \caption{12a shows the loading of the EEG files, and 12b and 12c are images generated from the DreamDiffusion [24].}
    \label{fig:side-by-side}
\end{figure}

\paragraph{Image Generation:}With the model, we could also improve functionality and generate images close to the original input (Figures 13 and 14).

\begin{figure}[h]
  \centering
  \includegraphics[width = 0.8\textwidth]{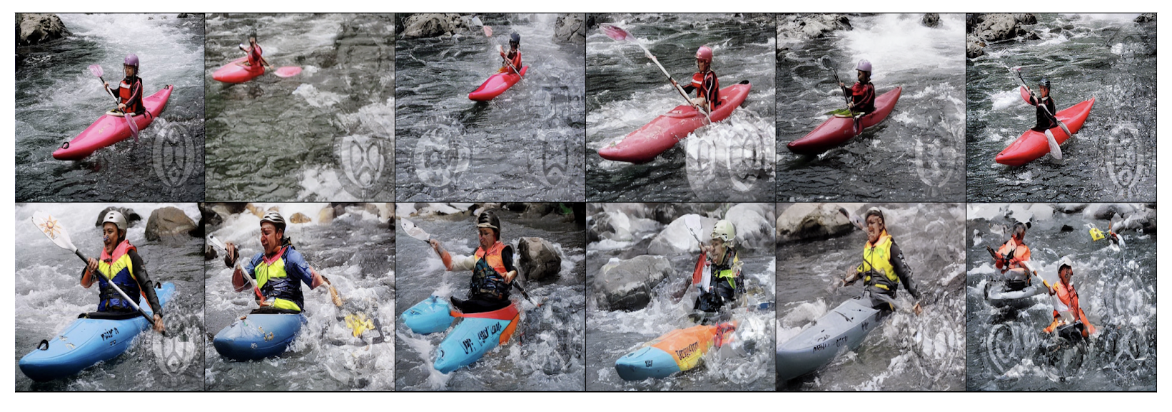}
  \label{fig:image}
  \caption{Shows images of a person kayaking [25].}
\end{figure}
\def\myvar{defined}

\begin{figure}[htbp]
    \centering
    \begin{minipage}[b]{0.45\textwidth}
        \centering
        \includegraphics[width=\textwidth]{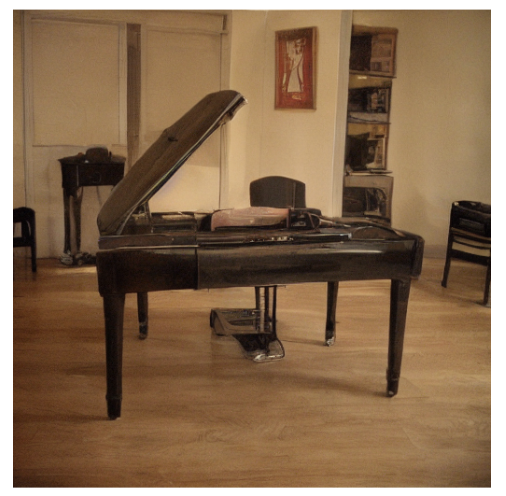}
    \end{minipage}
    \hspace{1cm} 
    \begin{minipage}[b]{0.45\textwidth}
        \centering
        \includegraphics[width=\textwidth]{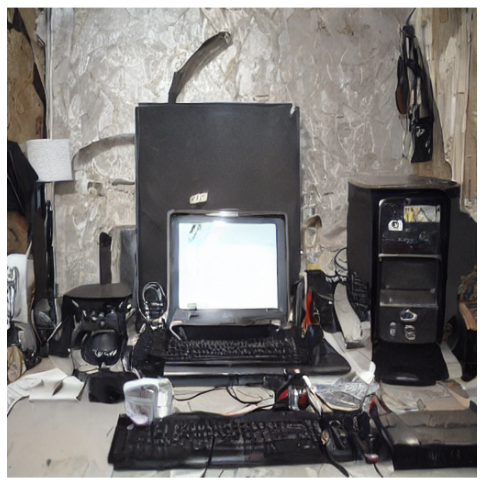}
    \end{minipage}
    \caption{Shows other images generated [26].}
    \label{fig:side-by-side}
\end{figure}

\section{Improvements}
\label{sec:headings}
\paragraph{Data Augmentation (DA) with Deep Learning:}Data augmentation techniques involve creating variations of existing data to increase the diversity of the dataset. This could include applying transformations such as time warping, amplitude scaling, or adding random noise to simulate real-world variability for EEG data. Deep learning methods can be employed to implement these augmentations effectively, ensuring that the generated data remains realistic and retains its underlying patterns.

\paragraph{Fourier Series for Wave-to-Time Conversion:}The Fourier series can decompose EEG signals into their constituent frequency components, allowing for a representation in the time domain. By performing Fourier analysis, we can identify dominant frequencies and their temporal evolution, providing insights into the rhythmic patterns of brain activity. Implementing Fourier series equations enables us to extract meaningful features from EEG data and accurately model temporal dynamics.

\paragraph{Sampling Ratio Optimization:}Optimizing the sampling ratio involves determining the appropriate proportions of data allocated for training and testing to maximize model performance. This optimization process requires careful consideration of dataset size, model complexity, and generalization requirements. We can find the optimal balance between training and testing data by experimenting with different sampling ratios to achieve superior model accuracy and robustness.

\paragraph{Nuclear Norm Regularized Deep Neural Network (NRDNN):}Nuclear norm regularization imposes constraints on the singular values of the weight matrices in deep neural networks, promoting sparsity and enhancing model generalization. By incorporating NRDNN into our model architecture, we can effectively differentiate various brain regions from EEG data and improve classification accuracy. This regularization technique helps prevent overfitting and enhances the stability and reliability of the trained model.

\paragraph{Activities of Daily Living (ADLs) Integration:}Integrating ADL-related EEG data into the model training process allows us to capture additional contextual information about the patient's state and environment. By incorporating features related to daily activities such as sleeping, eating, or exercising, the model can learn to recognize patterns associated with different states of consciousness and adapt its predictions accordingly. This integration enhances the model's ability to accurately assess the patient's condition and monitor their recovery progress.

\paragraph{Neural Network Techniques for Error Reduction:}Neural network techniques such as attention mechanisms, ensemble learning, and anomaly detection can help reduce human error in EEG data analysis by automating tedious tasks and flagging potential errors or inconsistencies. Attention mechanisms enable the model to focus on relevant features in the data, while ensemble learning combines multiple models to improve prediction accuracy. Anomaly detection algorithms can identify abnormal patterns in EEG signals, alerting clinicians to potential issues and facilitating timely intervention.

\paragraph{Decomposition Algorithms:}Decomposition algorithms such as principal component analysis (PCA), independent component analysis (ICA), and wavelet decomposition can extract meaningful features from complex EEG data, reducing dimensionality and improving model interpretability. PCA identifies orthogonal components that capture the most variance in the data, while ICA separates EEG signals into statistically independent sources. Wavelet decomposition analyzes EEG signals at different scales and resolutions, revealing temporal and frequency-domain information. By leveraging these decomposition techniques, we can enhance the model's ability to extract relevant information from EEG data and improve its performance in downstream tasks.

\paragraph{Bayesian Classification and Multi-class Common Spatial Patterns (MCSP):}Bayesian classification techniques provide probabilistic reasoning for decision-making, allowing for uncertainty quantification and robust model predictions. By incorporating Bayesian methods into our model, we can enhance its classification accuracy and reliability, particularly in scenarios with limited training data or noisy observations. MCSP techniques extract discriminative spatial patterns from EEG data, enabling the model to accurately differentiate between different brain states or conditions. Combining Bayesian classification and MCSP can improve the model's classification of EEG data and provide clinically relevant insights.

\paragraph{Morlet Wavelets for Power Estimation:}Morlet wavelets offer a powerful tool for time-frequency analysis of EEG signals, providing insights into the spectral characteristics of brain activity. By applying the Morlet wavelet transform to EEG data, we can estimate power spectra at different time points and frequencies, revealing patterns of neural oscillations associated with cognitive processes or neurological conditions. Morlet wavelets capture both temporal and frequency-domain information, making them well-suited for analyzing complex EEG data and improving the model's ability to detect subtle changes in brain dynamics over time.

\section{Existing Gaps in Research}
\label{sec:headings}
The main challenge during our project revolved around setting up the DreamDiffusion environment and seamlessly integrating all the necessary programs. We faced issues related to a lack of documentation, complex dependencies, and errors in code syntax, all of which demanded significant time and effort to resolve. As a collaborative effort, we have made the code readily available to facilitate further work on the model and encourage new implementations. However, there is a pressing need within the research community to streamline these resources, ensuring they are easily accessible and well-documented to minimize barriers to entry and facilitate smoother workflows.

Moreover, substantial gaps remain in efforts to enhance the quality of life (QoL) for patients with neurological conditions. With advancements in machine learning (ML) models and neural networks on the horizon, there is an increased urgency to leverage these technologies effectively in healthcare. Yet, there is a notable deficit in research focusing on leveraging these advancements to address patients' immediate needs and challenges. Researchers and practitioners must bridge this gap by developing and implementing innovative solutions that directly impact patient outcomes and improve their overall QoL. This entails a multidisciplinary approach incorporating input from healthcare professionals, technologists, and patients to ensure that emerging technologies are effectively translated into practical solutions that benefit those in need.

\section{Conclusion}
\label{sec:headings} In summary, traumatic brain injuries (TBIs) and associated neurological conditions represent critical areas of inquiry in contemporary medical research, holding profound implications for both patient outcomes and quality of life. Within this domain, the advent of innovative technologies such as DreamDiffusion offers promising avenues for advancing neuroimaging capabilities and therapeutic interventions. By harnessing pre-trained text-to-image models and employing sophisticated signal modeling techniques, DreamDiffusion endeavors to overcome inherent challenges in EEG data analysis. Through integration with CLIP image encoders and EEG signal encoders, DreamDiffusion seeks to align embedding spaces, enhancing the accuracy and efficacy of image generation from EEG signals. Nevertheless, the review underscores persistent gaps in research and implementation, particularly concerning the accessibility and usability of such advanced technologies. Complex setup procedures, dependency issues, and code errors remain significant impediments, necessitating concerted efforts to address and mitigate these barriers. Furthermore, there is a compelling imperative to prioritize enhancing patient quality of life through translating technological advancements into tangible clinical applications. Collaboration across multidisciplinary teams, including researchers, healthcare professionals, technologists, and patients, fosters continued research, innovation, and cooperation; the field can aspire to achieve transformative breakthroughs in diagnosis, treatment, and rehabilitation, ultimately improving outcomes and quality of life for individuals affected by TBIs and related neurological conditions. 

\section*{Acknowledgment Sample}
The completion of this dissertation would not have been possible without the help of Dr. Carlos Ponce (Harvard Medical School) and Dr. Suayb Arslan (Brain and Cognitive Sciences at MIT). They guided a group of high school students budding with interest in computational neuroscience through supplements, lectures, and assistance when problems arose. Specifically, work with progressing the model and optimization was with Dr. Ponce while editing and reviewing the literature was done on behalf of Dr. Arslan.

\clearpage
\bibliographystyle{unsrt}

\begin{thebibliography}{8}

\bibitem{nichd}
National Institute of Child Health and Human Development (NICHD). (n.d.). Traumatic Brain Injury: Condition Information. Retrieved from \url{https://www.nichd.nih.gov/health/topics/tbi/conditioninfo/treatment}

\bibitem{mayfield_clinic}
Mayfield Clinic. (n.d.). Patient Education: Traumatic Brain Injury. Retrieved from \url{https://mayfieldclinic.com/pe-tbi.htm}

\bibitem{ncbi}
National Center for Biotechnology Information. (2020). Neurological Manifestations of Hospitalized Patients with COVID-19 in Wuhan, China: A Retrospective Case Series Study. Retrieved from \url{https://www.ncbi.nlm.nih.gov/pmc/articles/PMC6952566/}

\bibitem{stanford_health}
Stanford Health Care. (n.d.). Coma Diagnosis. Retrieved from \url{https://stanfordhealthcare.org/medical-conditions/brain-and-nerves/coma/diagnosis.html#:~:text=Diagnosing%20a%20Coma,amount%20of%20consciousness%20and%20awareness.}

\bibitem{mayo_clinic}
Mayo Clinic. (n.d.). Electroencephalogram (EEG). Retrieved from \url{https://www.mayoclinic.org/tests-procedures/eeg/about/pac-20393875#:~:text=An%20electroencephalogram%20(EEG)%20is%20a,electrodes}

\bibitem{openai}
OpenAI. (n.d.). CLIP. Retrieved from \url{https://github.com/openai/CLI}

\bibitem{aws}
Amazon Web Services. (n.d.). Stable Diffusion. Retrieved from \url{https://aws.amazon.com/what-is/stable-diffusion}

\bibitem{arxiv}
Arxiv.org. (2023). DreamDiffusion: A Generative Model for EEG Representation. Retrieved from \url{https://arxiv.org/abs/2306.16934#:~:text=DreamDiffusion%20leverages%20pre%2Dtrained%20text,effective%20and%20robust%20EEG%20representations.}

\section*{Figures}

\bibitem{johns_hopkins}
Johns Hopkins Medicine. (n.d.). Anatomy of the Brain. Retrieved from \url{https://www.hopkinsmedicine.org/health/conditions-and-diseases/anatomy-of-the-brain}

\bibitem{mayfield_clinic}
Mayfield Clinic. (n.d.). Patient Education: Traumatic Brain Injury. Retrieved from \url{https://mayfieldclinic.com/pe-tbi.htm}

\bibitem{towards_data_science1}
Towards Data Science. (n.d.). Why is Logistic Regression the Spokesperson of Binomial Regression Models? Retrieved from \url{https://towardsdatascience.com/why-is-logistic-regression-the-spokesperson-of-binomial-regression-models-54a65a3f368e}

\bibitem{ncbi}
National Center for Biotechnology Information. (2020). Neurological Manifestations of Hospitalized Patients with COVID-19 in Wuhan, China: A Retrospective Case Series Study. Retrieved from \url{https://www.ncbi.nlm.nih.gov/pmc/articles/PMC6952566/}

\bibitem{stanford_health}
Stanford Health Care. (n.d.). Coma Diagnosis. Retrieved from \url{https://stanfordhealthcare.org/medical-conditions/brain-and-nerves/coma/diagnosis.html#:~:text=Diagnosing%20a%20Coma,amount%20of%20consciousness%20and%20awareness.}

\bibitem{grewal_2024}
Grewal, Fahim, Ellendula (2024).

\bibitem{fields}
Fields, R. D. (2018, December 3). Gamma Waves in the Brain: Fumes or Fundamental? Retrieved from \url{https://rdouglasfields.com/2018/12/03/gamma-waves-in-the-brain-fumes-or-fundamental/}

\bibitem{openai_clip}
OpenAI. (n.d.). CLIP. Retrieved from \url{https://github.com/openai/CLIP}

\bibitem{towards_data_science2}
Towards Data Science. (n.d.). What Are Stable Diffusion Models, and Why Are They a Step Forward for Image Generation? Retrieved from \url{https://towardsdatascience.com/what-are-stable-diffusion-models-and-why-are-they-a-step-forward-for-image-generation-aa1182801d46}

\bibitem{dream_diffusion}
Bbaaii. (n.d.). DreamDiffusion. Retrieved from \url{https://github.com/bbaaii/DreamDiffusion}

\bibitem{grewal_2024}
Grewal, Fahim, Ellendula (2024).

\bibitem{grewal_2024}
Grewal, Fahim, Ellendula (2024).

\bibitem{grewal_2024}
Grewal, Fahim, Ellendula (2024).

\bibitem{grewal_2024}
Grewal, Fahim, Ellendula (2024).

\end{thebibliography}

\end{document}